\documentclass[letterpaper]{article} 
\usepackage{marginnote} 
\usepackage{aaai25}  
\usepackage{times}  
\usepackage{helvet}  
\usepackage{courier}  
\usepackage[hyphens]{url}  
\usepackage{graphicx} 
\usepackage{dcolumn,booktabs}
\usepackage{natbib}  
\usepackage{caption} 
\usepackage{xcolor}
\newcommand{\answerYes}[1]{\textcolor{blue}{#1}} 
\newcommand{\answerNo}[1]{\textcolor{teal}{#1}} 
\newcommand{\answerNA}[1]{\textcolor{gray}{#1}} 
 
\usepackage{algorithm}
\usepackage{algorithmic}
\newcommand\mc[1]{\multicolumn{1}{c}{#1}}
\newcommand{\xhdr}[1]{\vspace{1mm}\noindent{{\bf #1.}}}

\usepackage{newfloat}
\usepackage{listings}
\DeclareCaptionStyle{ruled}{labelfont=normalfont,labelsep=colon,strut=off} 
\floatstyle{ruled}
\newfloat{listing}{tb}{lst}{}
\floatname{listing}{Listing}
\usepackage{cleveref}
\usepackage{xspace}
\usepackage{tabularx}
\usepackage{booktabs}
\usepackage{paralist} 
\usepackage{fontawesome5}
\usepackage{todonotes}
\usepackage{multirow}
\usepackage{arydshln}
\usepackage{chronology}

\title{Does Content Moderation Lead Users Away from Fringe Movements?\\Evidence from a Recovery Community}

\author {
    \footnote{Equal contribution}Giuseppe Russo\textsuperscript{\rm 1},
    $^*$Maciej Styzcen\textsuperscript{\rm 2},
    Manoel Horta Ribeiro\textsuperscript{\rm 3},
    Robert West\textsuperscript{\rm 1}
}
\affiliations {
    \textsuperscript{\rm 1}EPFL, \\
    \textsuperscript{\rm 2}Uber,\\
    \textsuperscript{\rm 3}Princeton University,  \\
    giuseppe.russo@epfl.ch, maciejs@uber.com, manoel@cs.princeton.edu, robert.west@eplf.ch\\
}

\usepackage{bibentry}

\begin{document}

\maketitle

\begin{abstract}
Online platforms have sanctioned individuals and communities associated with `fringe' movements linked to hate speech, violence, and terrorism---but can these sanctions contribute to the abandonment of these movements?
Here, we investigate this question through the lens of {r/exredpill}, a recovery community on Reddit meant to help individuals leave movements within the Manosphere, a conglomerate of fringe Web-based movements focused on men's issues.
We conduct an observational study on the impact of sanctioning some of Reddit's largest Manosphere communities on the activity levels and user influx of {r/exredpill}, the largest associated recovery subreddit. 
We find that {banning} a related radical community positively affects participation in {r/exredpill} in the period following the ban. Yet, \textit{quarantining} the community, a softer moderation intervention, yields no such effects. We show that the effect induced by banning a radical community is stronger than for some of the widely discussed real-world events related to the Manosphere and that moderation actions against the Manosphere do not cause a spike in toxicity or malicious activity in {r/exredpill}.
Overall, our findings suggest that content moderation acts as a deradicalization catalyst.
\end{abstract}

%

\section{Introduction} \label{section}

Users and communities associated with `fringe' movements like QAnon, Incel, or Proud Boys have been heavily sanctioned by mainstream social media platforms following their involvement with online harassment and real-world violence~\cite{bell2017reddit,qanonban2020,proudboys2018}.
Sanctions applied range from banning community and users permanently from the platform---`hard' content moderation~\cite{horta2021platform}---to reducing the visibility or flagging violations of community guidelines without removing them entirely---`soft' content moderation~\cite{zannettou2021won}.

\begin{figure}
    \centering
    \includegraphics[width=0.95\linewidth]{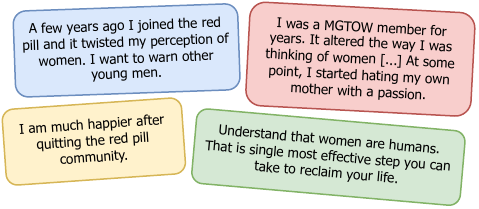}
    \caption{Some of the most upvoted comments and submissions in the \texttt{r/exredpill} recovery community (paraphrased due to privacy concerns).}
    \label{fig:exredpill-comments}
\end{figure}

While soft and hard moderation efforts are generally applauded by organizations that combat online violence and extremism~\cite{ADL2020,CCDH2023}, their effectiveness has been a subject of ongoing debate in academia~\cite{zuckerman2021deplatforming}.
On the one hand, moderation interventions have been shown to reduce the prevalence of hate speech and curtail activity in the targeted communities~\cite{chandrasekharan2017you, chandrasekharan2022quarantined}.
On the other hand, there are concerns about their unintended consequences; banned users often migrate to more radical, less regulated platforms, where their extremist views may intensify~\cite{horta2021platform} and spill over into mainstream platforms~\cite{russo2023spillover, russo2023understanding}.
However, this debate lacks evidence about \textit{recovery}. Do sanctions enacted by platforms lead users to increase their engagement with recovery communities? Does the extra work needed to find fringe content lead users to revisit their beliefs?

\vspace{1.5mm} \noindent \textbf{Present work.} 
In this paper, we address these very questions. We ask:

\begin{itemize} \item[] \textbf{RQ1}: Do \textit{soft} moderation interventions (e.g., quarantining) lead to increased participation in recovery communities?
\item[] \textbf{RQ2}: Do \textit{hard} moderation interventions (e.g., banning)
lead to increased participation in recovery communities? \end{itemize}
And finally, to compare the impact of real-world events on participation subsequent participation to the effects of moderation policies,  we ask:
\begin{itemize} \item[] \textbf{RQ3}: Do real-world riots and terrorist attacks lead to the participation in recovery communities?
\end{itemize}

We answer these research questions through the lens of recovery communities, online groups meant to foster user engagement with recovery communities--online spaces where individuals seek support to distance themselves from fringe ideologies.
While previous work has focused on aggregate trends of toxicity and participation in communities linked with fringe movements~\cite{make-reddit-great-again,horta2021platform,chandrasekharan2022quarantined}, we instead study how sanctions impacted participation in recovery communities, a metric that is more tightly linked with what may serve as an initial step toward disengagement from fringe ideologies (e.g., see Fig.~\ref{fig:exredpill-comments} for example comments).

We present a case study of r/exredpill, a large recovery that caters to individuals seeking to distance themselves from ideologies associated with the `The Manosphere'. The Manosphere is a conglomerate of anti-feminist movements~ (\textit{e.g.}, Incels, Men Going Their Own Way, Men's Rights Activists), all of which had large communities on Reddit, a mainstream social media platform~\cite{ribeiro2021evolution,manosphereMITTR, manosphereMisogyny}.
Manosphere-related communities on Reddit were repeatedly subjected to moderation interventions. Most notably, these communities have been \textit{banned}, a `hard' moderation measure that completely removes the community from Reddit, and \textit{quarantined}, a `soft' moderation measure that impedes direct access to and promotion of the community. 

We study the effect of quarantining (\textbf{RQ1}) and banning (\textbf{RQ2}) three Manosphere communities (r/MGTOW, r/Braincells, r/redpill) on three key participation-related outcomes on \textit{r/exredpill}: the overall activity within the support community, the influx of new participants, and the migration of users from fringe to recovery communities.
Further, we study the impact of three real-world events (Unite the Right Rally, Toronto Van Attack, Capitol Hill Siege) on the same participation-related outcomes (\textbf{RQ3}).
Our key analyses use two causal inference methods: interrupted time series (ITS) regression and Bayesian structural time series  (BSTS) modeling. 
Last, we conduct extensive robustness checks to ensure that increases in participation in \textit{r/exredpill} were not driven by negative comments or brigading users.

\vspace{1mm} \noindent \textbf{Results.} We found little evidence that soft moderation interventions (quarantining) increased participation in online recovery communities(\textbf{RQ1}). In contrast, we found that hard moderation interventions (banning) led to substantial increases in activity, newcomer participation, and migration to the recovery community r/exredpill. Following the bans of r/Braincels and r/MGTOW, activity levels increased by 88.4\% ($p=0.003$) and 64.5\% ($p=0.001$), respectively. Newcomer participation rose by 174.3\% ($p=0.004$) and 31.6\% ($p=0.001$), while migration to recovery communities grew by 94.6\% ($p=0.001$) and 22.8\% ($p=22.8$) after these bans (\textbf{RQ2}).
Surprisingly, while real-world events linked to these communities increased activity in the recovery community by up to 33\% and newcomer participation by 16\%, their impact was significantly smaller than that of platform-based moderation (\textbf{RQ3}).

\vspace{1mm} \noindent \textbf{Implications.} Mainstream platforms have attempted to mitigate the influence of fringe communities through visibility reduction and bans. However, these interventions are not without limitations, as fringe communities show remarkable resilience~\cite{horta2021platform, russo2024stranger}. Our findings suggest that these moderation actions might also serve as catalysts for deradicalization, potentially guiding users toward recovery communities. Platforms could consider leveraging this effect by strategically nudging fringe community members toward supportive environments, especially when impending moderation actions are anticipated.

\section{Background and Related Work}
\label{section:related-work}

\xhdr{Online Antisocial Communities}
 Fringe movements hold beliefs considered extreme by society at large~\cite{okholm2024debunking}, and online communities associated with said movements are known to frequently engage in antisocial behavior, propagate conspiracy theories, and promote extremist ideologies~\cite{marwick2018drinking, russo12024automatic}.
Examples of these movements include QAnon~\cite{schulze2022far}, the Alt-right~\cite{rieger2021assessing}, and most relevant to the work at hand, the Manosphere, a conglomerate of anti-feminist movements characterized by their hostility towards women~\cite{manosphereMisogyny}.
Manosphere-related communities prospered on Reddit in the 2010s, in `subreddits' [discussion forums; see ~\citet{ribeiro2021evolution} for details].
Here, we focus on three Manosphere communities active on Reddit, each associated with a different movement within the Manosphere: {r/Braincels}, {r/MGTOW}, and {r/TheRedPill}. In the paragraphs below, we briefly describe the communities and their associated movements.

 {r/Braincels} was a subreddit associated with the Involuntary Celibate movement. 
 The online community became popular circa 2017 after another subreddit (r/Incels) was banned by Reddit for breaching community guidelines~\cite{baele2024diachronic}. The Incel movement abides by ``The Black Pill,'' the idea that some men (Incels) are unable to have romantic and sexual relationships because of their physical appearance~\cite{incelswiki}.
Incel communities are notorious for creating hateful and misogynistic content~\cite{ribeiro2021evolution}.
Further, Incels believe that systemic changes are needed to address men's dating issues and have supported and perpetrated acts of violence to achieve them~\cite{o2022political}.

{r/MGTOW} was a subreddit associated with the Men Going Their Own Way movement. It was created in 2011, and was among the most popular Manosphere-related subreddits~\cite{ribeiro2021evolution}. The Men Going Their Own Way movement preaches that society is rigged against men~\cite{Lin+2017+77+96} and that the only solution is the abandonment of women and,  sometimes, of western society in general~\cite{Lin+2017+77+96, jones2020sluts}.
Members of the movement openly disdain women, and MGTOW-adjacent online communities propagate and normalize misogynistic beliefs via online harassment~\cite{jones2020sluts}

{r/TheRedPill} is a subreddit associated with various movements within the Manosphere founded by a New Hampshire state legislator in 2012~\cite{trpstate}.
The subreddit's name alludes to a famous scene from the movie ``The Matrix''  that, within the Manosphere, refers to the (internally widespread) belief that men, and not women, are disadvantaged in modern (feminist) society~\cite{ging2019alphas}.
Much of r/TheRedPill content describes pseudo-scientific `sexual strategy in a culture increasingly lacking a positive identity for men,' alluding to Manosphere movements like Pick Up Artists, Men's Rights Activists, and Men Going Their Own Way~\cite{thorburn2023recovery2, ribeiro2021evolution}.
Perhaps unsurprisingly, r/TheRedPill was notorious as a hub for misogynistic content on Reddit~\cite{redpill}.

\xhdr{Impact of community-level sanctions} Incidents of online harassment, hate speech, and real-world violence led Reddit to sanction communities associated with fringe movements. Typically, Reddit has applied one of two sanctions: quarantining (soft moderation) and banning (hard moderation). 
Quarantined subreddits do not appear on user's feeds, are not included in search or recommendations, require users to be logged in to Reddit to view the community, and display a warning that requires users to explicitly opt-in to view the content~\cite{quarantined}.
Banned subreddits are deleted from Reddit, and all their posts and comments become inaccessible. Notably, users participating in a banned subreddit keep their accounts.
Prior research shows that quarantines reduce new user recruitment, though the effect is often modest~\cite{chandrasekharan2017you, trujillo2022make},  but do not significantly reduce existing users' toxicity~\cite{chandrasekharan2022quarantined}. Bans significantly reduce activity but can push users to other fringe platforms~\cite{horta2021platform}, leading to spillover effects back onto mainstream platforms~\cite{russo2023spillover,schmitz2022quantifying, russo2024shock}.

\xhdr{Recovery communities} Recovery communities on Reddit support individuals dealing with issues like addiction~\cite{recovery-alcohol,recovery-opiod, recovery-opioid-help}, mental health problems~\cite{recovery-mental}, eating disorders~\cite{recovery-eating-disorders}, and even political extremism and conspiracy beliefs~\cite{recovery-qanon,recovery-qanon-2}. These communities provide networks to aid in recovery or deradicalization.
This study focuses on r/exredpill, a key recovery community for members of the Manosphere, including subreddits like r/Braincels, r/MGTOW, and r/TheRedPill. While previous research highlights r/exredpill's potential for de-radicalization~\cite{thorburn2023recovery1, thorburn2023recovery2, incelexit-qualitative}, it has primarily relied on qualitative analysis, with no quantitative studies conducted yet.

Research on the recovery from extremist beliefs often relies on interview-based studies and theoretical models of disengagement. For instance, \citet{xiao2021sensemaking}  conducted interviews with current and former chemtrail conspiracy believers, revealing that accidental exposure to counter-narratives \cite{engel2023learning}, persuasion by trusted peers, and a desire for social acceptance were key factors in abandoning such beliefs. 
Similarly, studies on QAnon communities \cite{jigsaw2021conspiracy, phadke2021characterizing} have used qualitative methods to highlight the role of disillusionment with failed predictions and unmet promises in prompting recovery.

Theoretical models further frame these processes. Cognitive Dissonance Theory ~\cite{harmon2019introduction} suggests that exposure to conflicting information or the need to express opposing views publicly can induce internal tension,  leading to attitude shifts. Role Exit Theory ~\cite{ebaugh1988becoming} describes disengagement as a staged process involving 
1) doubt, 
2) exploration of alternatives, 
3) a decisive turning point, and
4) the establishment of a new identity. 
~\citet{aho1988out} identifies various pathways toward radicalization and deradicalization, but most importantly to the work at hand, characterizes voluntary radicalization and deradicalization as a consequence of a change in the push and pull factors.


\xhdr{Present and prior work} 
Our hypotheses draw on deradicalization theories, such as role exit~\cite{ebaugh1988becoming} and Aho's Defection Model~\cite{aho1988out}. Both theories frame deradicalization as a multi-step process, with a key step linked to external events that can exacerbate and accelerate preexisting doubts, ultimately leading to the abandonment of the community. We hypothesize that moderation policies, like quarantines and bans, may act as ``turning points'' by disrupting engagement, fostering cognitive dissonance, and prompting belief reassessment~\cite{harmon2012cognitive, wacquant1990review}. Quarantines might expose users to alternative perspectives, enabling gradual disengagement~\cite{xiao2021sensemaking}, whereas bans may trigger abrupt disruptions but risk reinforcing oppositional identities~\cite{liguori2021recovering, berube2019converging}. 


\section{Materials and Methods}\label{sec:data}

\vspace{1.25mm}

\subsection{Data Collection}

We used the Reddit Archive (\url{the-eye.eu}) to retrieve all comments and posts from three prominent fringe subreddits: r/Braincels, r/TheRedPill, and r/MGTOW.
We considered all comments and posts from their creation until their eventual banning from Reddit. In the case of r/TheRedPill, which was only quarantined, we collected comments until 180 days after the quarantine event.
Consistent with prior work, we considered only comments from users who contributed with more than five comments or posts to any of these subreddits~\cite{kumar2018mega, samory2018conspiracies}. Following \citet{russo2024stranger}, if a user exceeded the threshold across multiple subreddits, we categorized them under the subreddit with the highest activity to avoid duplicate classifications across communities. Our dataset comprises 9.8 million comments and 574,057 submissions from these communities, with further details presented in \Cref{tab:data}.

\begin{table*}[ht]

\label{tab:data_summary}

\centering
\footnotesize
\begin{tabular}{lccccc}
\toprule

 & Comments & Submissions & Users &  Quarantine  &  Banning \\
\midrule
\emph{Fringe subreddits} &  &  &   &   &  \\ \midrule

r/Braincels & $2{,}826{,}336$ & $216{,}806$ & $50{,}379$ & \textcolor{green}{\faCheckCircle} & \textcolor{green}{\faCheckCircle} \\
r/MGTOW & $5{,}449{,}655$ & $290{,}503$ & $122{,}526$ & \textcolor{green}{\faCheckCircle} & \textcolor{green}{\faCheckCircle} \\
r/TheRedPill & $3{,}206{,}546$ & $118{,}396$ & $122{,}684$ & \textcolor{green}{\faCheckCircle} & \textcolor{red}{\faTimesCircle} \\
\midrule
\emph{Recovery subreddits} & & & & & \\\midrule
r/exredpill & $176{,}035$ & $8{,}221$ & $12{,}720$ & --- & --- \\
\bottomrule
\end{tabular}
\caption{For the subreddits considered in this paper, we depict the number of comments, submissions, and users obtained (columns 1--3) via the data collection of the entire posting (comments+submissions) history. Also, for each subreddit, we include information about the kind of moderation the community received (i.e., quarantine or banning; columns 4--5).}
\label{tab:data}

\end{table*}

The subreddits studied were subject to various moderation actions, including quarantines and bans. Drawing on media reports~\cite{mgtow-ban-news, braincels-ban-news} and documentation from the r/reclassified subreddit (which documented Reddit sanctions), we identified four key sanctions applied to r/Braincels, r/TheRedPill, and r/MGTOW:
\begin{enumerate}
    \item On September 27th, 2018,  r/Braincels and r/TheRedPill were quarantined after Reddit updated its policies on content moderation;
    \item On October 1st, 2019, r/Braincels was banned;
    \item On January 31st, 2020, r/MGTOW was quarantined;
    \item On August 3rd, 2021, r/MGTOW was banned.

\end{enumerate}


Finally, we collected all data of r/exredpill, a recovery community offering peer support to those disengaging from the ideologies promoted in r/Braincels, r/TheRedPill, and r/MGTOW. We gathered in total 8{,}221 submissions and 176{,}035 comments from 12{,}720 users made within a 120-day window before and after each identified moderation event. We defined membership in r/exredpill based on users who posted at least five times within the subreddit. 

To accurately measure the effects of moderation events on the subsequent particpation in recovery community, we operationalize participation via three outcome variables:

\vspace{1mm} \begin{enumerate} 
\item \textbf{Activity Volume:} The daily number of comments and submissions in r/exredpill. 
\item \textbf{Number of New Users:} The number of users posting in r/exredpill for the first time on a given day.
\item \textbf{Migrating Users:} The number of users posting in r/exredpill for the first time after previously contributing to one of the radical manosphere communities. \end{enumerate} \vspace{1mm}

\vspace{1.25mm}

\subsection{Estimating the Causal Effect}\label{sec:causal_effects}

\vspace{1.5mm}
\noindent
To estimate the causal effects of soft and hard moderation interventions (\textbf{RQ1} and \textbf{RQ2}), as well as of real-world events (\textbf{RQ3}), we use two causal inference methods: interrupted time series (ITS) regression and Bayesian structural time series (BSTS) modeling.

\vspace{1.5mm}
\noindent
\textbf{Interrupted time series analysis (ITS)} is a widely used technique for detecting changes in trends, onset, and decay of effects from interventions by examining a series of observations before and after a defined intervention point~\cite{bernal2017interrupted}.
The applicability of ITS depends on certain data assumptions. For example, when non-linear trends or specific distributions are present, more advanced regression techniques may be required~\cite{wagner2002its}. 
ITS models must address issues like autocorrelation and seasonality, which can skew effect size estimates if not properly accounted for. In our study, we utilize the ITS regression model fo to illustrate changes in linear trends of key variables around the intervention points, as described in the linear model below:

\begin{equation}
    Y_t = \beta_0 + \beta_1 T + \beta_2 D + \beta_3 P + \epsilon,
    \label{eq:its}
\end{equation}
where, $Y_t$ represents the outcome variable of the time series, $T$ is a continuous variable indicating time in days from the start of the observational period, with $\beta_1$ capturing the pre-intervention trend. $D$ is a binary variable indicating the presence (1) or absence (0) of the intervention, with $\beta_2$ representing the immediate effect of the intervention. $P$ is a continuous variable indicating the number of days since the intervention, with $\beta_3$ capturing any post-intervention trend changes. Finally, $\epsilon$ represents the model's error term.

The ITS model was fit using Ordinary Least Squares (OLS), chosen for its simplicity and suitability for visualization rather than inferential purposes. While count outcome variables may often be skewed, we prioritized OLS to emphasize absolute changes, aligning with our goal of identifying and visualizing trends. Additionally, this approach served as a robustness check complementary to the Bayesian Structural Time Series (BSTS) model.

\vspace{1.0mm}
\noindent
\textbf{Bayesian Structural Time Series (BSTS) Modeling} is a Bayesian statistical approach that offers several advantages over traditional ITS analysis. BSTS allows for the decomposition of a time series into components, combined with a dynamic regression framework that uses Monte Carlo Markov Chain (MCMC) simulations to generate counterfactual data and confidence intervals~\cite{brodersen2015inferring}.

The method estimates a synthetic control via a state-space time-series model that uses information from 1) the time-series behavior of the outcomes of interest and 2) a set of multiple control time series similar to the target series.  The synthetic control is made on the pre-treatment portion of potential controls, but its value lies in the post-treatment period. As long as the control series received no intervention, it is reasonable to assume the relationship between the treatment and the control series that existed before the intervention will continue afterward. Thus, a plausible estimate of the effect of the intervention can be computed.

To identify such control time series, we select subreddits that share demographic and political characteristics with r/Braincels, r/TheRedPill, and r/MGTOW. These control subreddits predominantly feature young male users with right-leaning views. To ensure comparability, we assessed subreddits across three social dimensions—partisanship, age, and gender—using cosine similarity to match them with the treatment subreddits (using social dimensions provided by ~\cite{waller2021quantifying}. The final control group includes 48 subreddits such as \texttt{r/Conservative}, \texttt{r/cigars}, \texttt{r/GunPorn}, and \texttt{r/mancave}.

BSTS is a more robust method for estimating intervention effects, particularly in the presence of autocorrelation and seasonality in the data. For our analysis, we utilize the BSTS implementation provided by the CausalImpact R package with MCMC 1000 iterations to ensure robust inference of the intervention effects.

\section{Results}

\begin{figure*}[t]
    \centering
\includegraphics[width=0.7\textwidth]{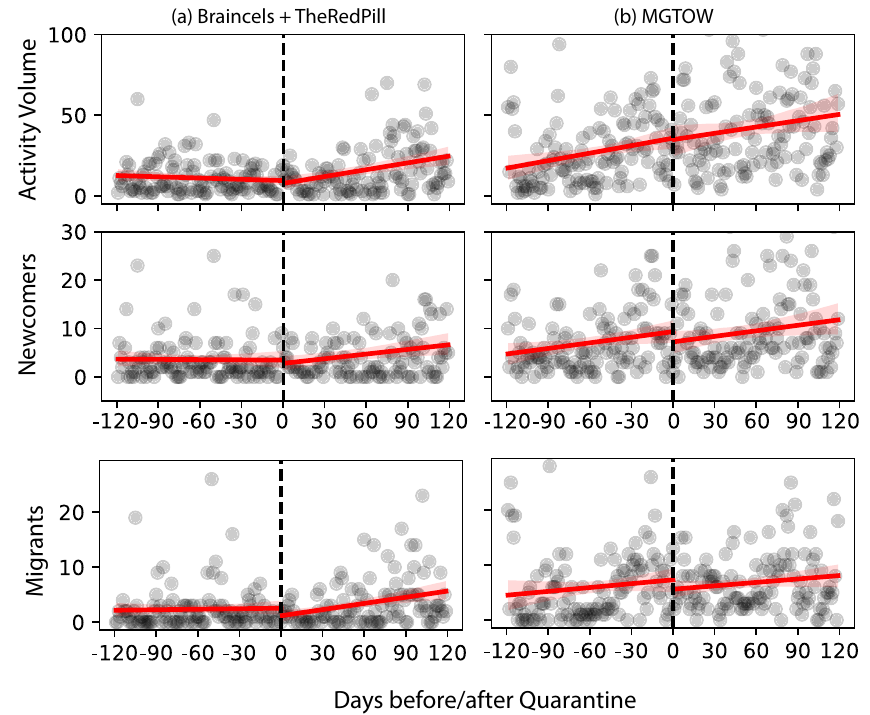}
    \caption{{Effects of Quarantine on r/exredpill Activity Volume, Newcomers, and Migrants.} We show the results obtained after fitting the ITS model on the activity volume, newcomers, and migrants that joined r/exredpill in the aftermath of the quarantine. 
    In the left column (a), we show the ITS analysis assessing the effect on activity volume and newcomers after the quarantine of r/Braincels and r/TheRedPill.
    In right column (b), we show the same analysis after the quarantine of r/MGTOW.
    In the bottom row, we show the ITS analysis for the users who joined r/exredpill after having participated in r/Braincels (a), r/TheRedPill (b), and r/MGTOW (c). 
    }
    \label{fig:quarantine}
\end{figure*}

\begin{figure*}[t]
    \centering
\includegraphics[width=0.65\textwidth]{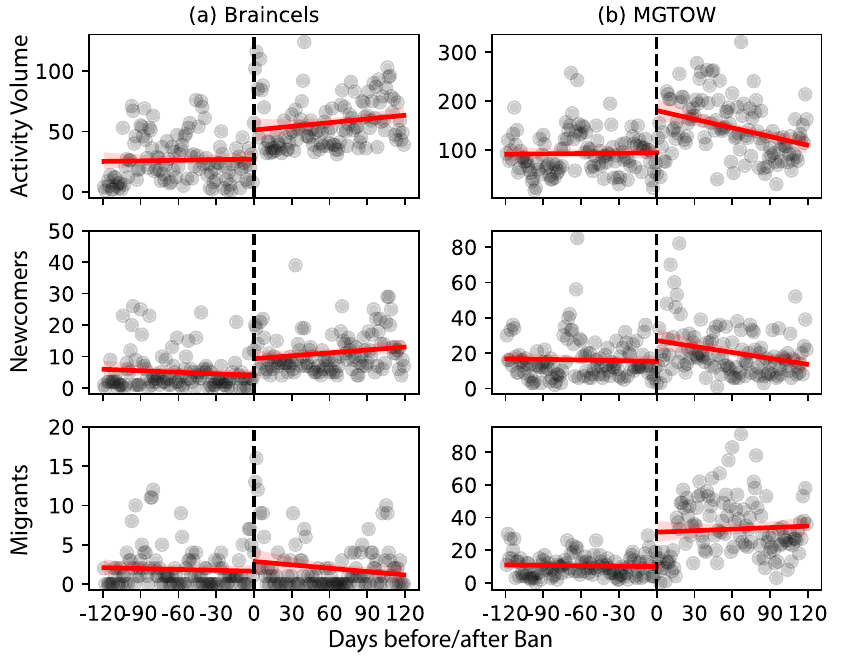}
    \caption{{Effects of Banning on r/exredpill Activity Volume, Newcomers, and Migrants.} We show the results obtained after fitting the ITS model on the activity volume, newcomers, and migrants that joined r/exredpill in the aftermath of the banning. 
    In the left column (a), we show the ITS analysis assessing the effect on activity volume and newcomers after the ban of r/Braincels.
    In right column (b), we show the same analysis after the banning of r/MGTOW.
    }
    \label{fig:banning}
\end{figure*}

We examine the effect of quarantining, banning, and real-world events on participation-related outcomes associated with r/exredpill. We consider two quarantining events (the quarantining of r/Braincels and r/TheRedPill in 2018, and   of r/MGTOW in 2020),  two banning events (the banning of r/Braincels and r/MGTOW), and three real-world events (Unite the Right Rally, Toronto Van Attack, Capitol Hill Siege).
The ITS analysis and the BSTS modeling results are summarized in \Cref{table:coefficients,table:banning-results,table:quarantine-results} (at the end of this document). We show the time series of the key outcomes and the estimated regression lines in \Cref{fig:quarantine,fig:banning}.

\subsection{RQ1: Effect of quarantining}

\xhdr{Quarantining and Activity Volume}
Using Interrupted Time Series (ITS) analysis, we found no significant immediate change after the event ($\beta_2$ in the model), but we did observe varying impacts on activity trends following these events.
The trend in activity in r/exredpill increased following the quarantining of r/Braincels and r/TheRedPill ($\beta_{3}^{BI+TRP} = 0.16; p = 0.002$), suggesting these events may have sparked discussion or attracted new users. Yet, we find no significant increase in activity following the quarantining of  r/MGTOW.
Since ITS lacks a control group, platform-wide trends could have influenced the results. To account for this, we applied BSTS modeling. This subsequent analysis casts doubt on the validity of the ITS results, as we find non-significant increases following the quarantining of r/Braincels and r/TheRedPill (15.4\%; $p = 0.345$) and of r/exredpill (38.5\%; $p = 0.09$).
In this context, we conclude that quarantining fringe subreddits did not significantly impact activity volume in r/exredpill.

\xhdr{Quarantining and Number of New Users}
Next, we examine whether quarantining influences the influx of new users to r/exredpill. 
In both the ITS and the BSTS analyses, we find no statistically significant effect of the moderation intervention on the number of newcomers, suggesting that quarantine did not help popularize r/exredpill.

\xhdr{Quarantining and Migrating Users}
Finally, we explore whether quarantining fringe communities spurred migration to r/exredpill from r/Braincels,  r/TheRedPill, and r/MGTOW.
In the ITS analysis,  we find no significant effect on the number of users that migrated from r/Braincels and r/TheRedPill to r/exredpill immediately after the quarantine ($\beta_2=-0.01; p=0.318$). 
Differently, we observe that quarantining led to a significant increase in the trend of migrants that previously participated in r/Braincels or r/TheRedPill ($\beta_3^{BI+TRP} = 0.044; p = 0.02$), suggesting that quarantining these subreddits led to an uptick in user migration to the recovery community in the period following up the quarantine.
These results are confirmed by the BST analysis, which shows a 24.6\% ($p = 0.045$) increase in migration from r/Braincels and r/TheRedPill to r/exredpill. In contrast, we found no statistically significant changes in the number of migrants from r/MGTOW to r/exredpill in the aftermath of the r/MGTOW quarantine (see \Cref{table:quarantine-results}).
Altogether, these different results support the notion that quarantining increases the number of migrating users. However, we argue they do not provide substantial evidence that soft moderation interventions increase participation in recovery communities. Given that we are considering three metrics and two different events, this is likely a spurious finding. For instance, a conservative Bonferroni correction would set the significance threshold at 0.0083 (0.05$\div$6), rendering the effects observed here not statistically significant.


\subsection{RQ2: Effect of banning}

\xhdr{Banning and Activity Volume}
Upon the banning of both r/Braincels and r/MGTOW, we observe a statistically significant increase in the activity volume in both the ITS analysis ($\beta_2^{BI}=24.06; p<0.001$, $\beta_2^{MT}=86.49; p<0.001$) and in the BSTS model (88.4\% and 64.5\% increase; $p=0.032$ and $p=0.001$).  
After the banning of r/Braincels, the activity volume of r/exredpill remains similar to the activity level at the time of the banning ($t=0$). The coefficient $\beta_3$ that captures the trend growth  after the banning ($t>0$)  is positive (indicating a slightly increasing trend) but not statistically significant.
After the banning of r/MGTOW, differently from the r/Braincels ban, we observe a statistically significant decreasing trend ($\beta_3=-0.61;p=0.001$) in activity volume. However, even if the trend decreases, the activity volume remains consistently higher than the pre-ban activity volume. 
These results indicate that the banning of Manosphere communities increased activity in r/exredpill.

\xhdr{Banning and Newcomers} Considering the influx of newcomers, we observe that the banning events influenced the number of users joining r/exredpill. The number of new users who posted in the recovery community rose significantly following the bans of r/Braincels and r/MGTOW ($\beta_2^{BI}=5.30; p=0.001, \beta_2^{MT}=11.74; p<0.000$).
The number of newcomers in the days following the ban exhibits an increasing statistically significant linear trend for r/Braincels ($\beta_2^{BI}=0.04; p=0.04$) and a not statistically significant decreasing trend for r/MGTOW.
The BSTS analysis further supports the ITS, showing an increase in the number of newcomers following the r/Braincel and r/MGTOW ban of 174.3\% ($p=0.004$) and 31.7\% ($p=0.001$), respectively.

\xhdr{Banning and Migrating Users} Finally, we examine the migration of users from the banned fringe communities to r/exredpill. Following the bans of r/Braincels and r/MGTOW, we observed an uptick in the number of users from these communities posting in the recovery subreddit. The ITS analysis highlights a positive and significant increase in migration to r/exredpill after the r/Braincels and ($\beta_2^{BI}=1.24; p=0.03$) and r/MGTOW banning ($\beta_2^{MT}=21.10; p<0.001$).
Similarly to  what we observed in the case of Activity Volume and Newcomers, we observe one slightly decreasing linear trend after the Braincels ban ($\beta_3^{BI}=-0.01; p=0.318$), and a increasing linear trend after the ban of MGTOW ($\beta_3^{MT}=0.04;p=0.441$). However, neither change in trend is statistically significant. The BSTS analysis estimates a 94,6\% ($p=0.001$) increase in migrating users from r/Braincels  corresponding to an absolute increase of 46 users, and a 22.8\% increase from r/MGTOW ($p = 0.006$), 21 additional users. These findings suggest that banning fringe communities not only curtails their activity but also prompts a subset of users to seek out recovery and support.

\begin{table*}[t]
\small
\centering
    \begin{tabular}{l|r:r:r}
    \textbf{Event}                       & p-value & relative effect & 95\% CI            \\ \midrule
    \textbf{Braincels ban}               & 0.105   & 82\%            & {[}-13\%, 250\%{]} \\
    \textbf{MGTOW ban }                  &\textbf{0.001}   & -55\%           & {[}-63\%, -44\%]    \\
    \textbf{Braincels \& TRP quarant.} & 0.279   & -0.57\%         & {[}-46\%, 155\%{]} \\
    \textbf{MGTOW quarantine }           & 0.362   & 9\%             & {[}-20\%, 56\%{]} \\ \bottomrule
    \end{tabular}
    \caption{Effect of content moderation events against 'Manosphere' communities on the fraction of comments with a deleted body in \texttt{r/exredpill}.}
    \label{table:deleted}
\end{table*}

\subsection{RQ3: Comparison with Real-World Events}
To assess the impact of real-world events connected to the Manosphere on participation in recovery communities, we analyzed three significant events: the Unite the Right rally (UR) on August 11, 2017, the Toronto Van Attack (TA) in April 23, 2018, and the Capitol Hill Siege (CH) in January 6, 2021. These events were chosen for their ideological ties to the Manosphere. 
A self-identified Incel perpetrated the Toronto Van Attack~\cite{TheGuardian_2019}, and both the Unite the Right Rally and the Capitol Hill Siege were associated with far-right groups tightly associated with the Manosphere~\cite{mamie2021anti}.

We first examined their effect on activity volume within r/exredpill. The BSTS analysis revealed statistically significant increases in activity following all three events, with activity volume rising by 11.2\% ($p = 0.012$), 22.1\% ($p = 0.06$), and 33.5\% ($p = 0.043$) for the Unite the Right rally, the Toronto Van Attack, and the Capitol Hill Siege, respectively. The Interrupted Time Series (ITS) analysis similarly detected increases in activity immediately following these events ($\beta_2^{UR} = 2.82$; $p = 0.529$, $\beta_2^{TA} = 4.20$; $p = 0.018$, $\beta_2^{CH} = 11.65$; $p = 0.006$). However, no statistically significant changes in trends were observed after the events. Notably, compared to the effects of bans on r/MGTOW (+64.5\%) and r/Braincels (+88.4\%), the activity increases following real-world events were much smaller.
Next, we analyzed the influx of new users into r/exredpill. The BSTS analysis found a statistically significant increase (+19\%, $p = 0.012$) in new users following the Capitol Hill Siege. Still, no significant effects on newcomer influx were identified following the Unite the Right rally or the Toronto Van Attack. In contrast, the ITS analysis did not reveal statistically significant changes in newcomers immediately following ($\beta_2$) or after ($\beta_3$) any of the events.
Finally, considering the migration of users from fringe communities to r/exredpill, neither the ITS nor the BSTS analyses found any statistically significant changes in migration patterns following these events.
Altogether, these results provide some evidence that real-world riots and terrorist attacks boost participation to recovery communities. Yet, most important for the work at hand, they highlight the magnitude of the effect sizes observed as a result of hard content moderation interventions.

\begin{table*}[]
\centering
\renewcommand{\arraystretch}{1.1}
\begin{tabular}{c|l|r:r:r}
\textbf{Event} &\textbf{Outcome variable} & \textbf{p-value} & \textbf{relative effect} & \textbf{95\% CI}           \\ \toprule
\multirow{3}{*}{\textbf{Unite the Right Rally (UR)}}  
& activity volume  & \textbf{0.012}   & \textbf{11.2\%}           & \textbf{{[}6.1\%, 16.3\%{]}}  \\
& new users        & 0.645   & 0.8\%            & {[}-2.3\%, 3.9\%{]}  \\
& migrating users  & 0.732   & -1.1\%           & {[}-3.5\%, 1.3\%{]}  \\ \hdashline
\multirow{3}{*}{\textbf{Toronto Van Attack (TA)}}  
& activity volume  & 0.06    & 22.1\%           & {[}15.1\%, 29.1\%{]}  \\
& new users        & 0.559   & -0.6\%           & {[}-2.7\%, 1.5\%{]}  \\
& migrating users  & 0.687   & 1.2\%            & {[}-1.8\%, 4.2\%{]}  \\ \hdashline
\multirow{3}{*}{\textbf{Capitol Hill Siege (CH)}}  
& activity volume  & \textbf{0.043}   & \textbf{33.5\%}           & \textbf{{[}26.3\%, 40.7\%{]}} \\
& new users        & \textbf{0.012}   & \textbf{19.0\%}           & \textbf{{[}12.3\%, 25.7\%{]}}  \\
& migrating users  & 0.831   & -0.3\%           & {[}-2.6\%, 2.0\%{]}  \\ \bottomrule
\end{tabular}
\vspace{10pt}
\caption{BSTS results of the effects of real-world events on volume activity, newcomers, and migrating users within r/exredpill.}
\label{table:real-world-events-results}
\end{table*}

\section{Robustness Checks}

A critical aspect of our analysis is ensuring that the observed increase in activity within r/exredpill following the moderation of fringe communities is not driven by negative motivations, such as brigading from users of the sanctioned communities. To address this concern, we conducted a series of robustness checks focusing on the content of the comments posted in r/exredpill after these moderation events.

\xhdr{Deleted content}
We examined the fraction of deleted comments and submissions in r/exredpill before and after the moderation events. If the increased participation in r/exredpill was driven by former users brigading against r/exredpill, we would expect to see a rise in deleted content following the interventions.
Yet, our analysis found no significant changes in the deleted content fraction in three of the four events studied. The only exception was a decrease observed following the ban of r/MGTOW, which we attribute to an upward trend in deleted comments during the pre-intervention period rather than the intervention itself. Overall, these findings suggest that the increase in activity within r/exredpill did not lead to a rise in inappropriate or rule-violating content. We provide the results of this analysis in \Cref{table:deleted}.

\xhdr{Toxicity Analysis} We investigated whether the increased activity in r/exredpill could be attributed to an influx of toxic behavior from users of the moderated communities. To do this, we analyzed the toxicity levels of comments posted before and after each moderation event. We used Google's Perspective API~\cite{perspective} to annotate all comments with a toxicity score and applied Interrupted Time Series (ITS) and Bayesian Structural Time Series (BSTS) models to detect significant changes.
Our analysis found no statistically significant increases in average toxicity levels of r/exredpill around the time of the moderation events. Specifically, the ITS models showed no immediate change in toxicity at the time of intervention or in the follow-up period. Similarly, the BSTS models did not indicate any significant shifts in toxicity levels. These results held for all the moderation events considered.

\xhdr{LLM Moderation Analysis} 
While toxicity offers a useful indicator for assessing whether increased participation was driven by retaliation from moderated communities, it remains a controversial measure for content moderation~\cite{bias_perspective2023, gargee2023toxicity}. Inspired by previous works that used Large Language Models (LLM) to annotate data \cite{latona2024ai, davidson2024self}, we used GPT-4-turbo, a large language model, to further evaluate the nature of comments posted by newcomers and migrants in r/exredpill.
We applied two strategies. First, we provided GPT-4-turbo with the community guidelines of r/exredpill and a comment posted on r/exredpill from migrating users, asking if it violated any rules. Second, we described the ideologies of the three fringe communities (r/Braincels, r/MGTOW, and r/TheRedPill) and asked if the comment aligned with those ideologies.
GPT-4-turbo labeled  97\% of the comments as compliant with the  community guidelines, and 99\% did not align with the values of the fringe communities. This analysis reinforces that the post-moderation activity in r/exredpill reflects recovery rather than a continuation of fringe ideologies.
We provide the prompts used in the Appendix.

\xhdr{Observation Window}To ensure the robustness of our findings, we tested various observation window lengths, including 60, 90, 150, and 180 days. Our sensitivity analysis showed that the results remained consistent across these different windows. This consistency suggests that our findings are not influenced by the choice of observation window. We selected the 120-day window as it provides a practical balance between capturing relevant trends and maintaining temporal proximity to the events studied.
 
\xhdr{Placebo Testing} To evaluate the reliability of our results, we conducted placebo tests using Bayesian Structural Time Series (BSTS) modeling. Specifically, we introduced “placebo intervention” dates at -120 days and +120 days relative to each studied intervention. We report these results in \Cref{table:banning-results-placebo,table:quarantine-results-placebo} These placebo tests were applied to all quarantine and banning events where our main analysis detected changes. Additionally, we repeated this process using alternative observation windows of 60, 90, 150, and 180 days.
Across all scenarios, the results consistently showed no effects for the placebo intervention dates.

\xhdr{Manual Inspection} To further understand the content of the posts published by either newcomers or migrants, we go beyond the automatic content analysis and perform a human judgment analysis.
We selected 200 random comments made by users who are either newcomers or migrants from the r/exredpill recovery community from one of the r/Braincels, r/MGTOW, and r/TheRedPill communities after the moderation action had been taken. Two human annotators, both authors of this paper, labeled these sentences by marking whether they contain pro-Manosphere content or attacks against the r/exredpill community.
In addition, we report interannotator rates with Cohen’s $\kappa$. In 94\% of the comments, no pro-Manosphere or community attack was identified. Instead, many comments reflected a change in views from those previously held by users in the fringe communities.

\section{Discussion}

Our study investigates the effects of moderation policies on fringe communities and their potential to influence participation in recovery communities. We shed light on how these interventions shape user behavior and recovery processes by examining soft  (quarantines) and hard moderation (bans) interventions alongside real-world events associated with fringe ideologies. 

Our key findings are threefold. First, we find that, contrary to our initial hypothesis (\textbf{RQ1}), quarantines of fringe communities had no substantial impact on recovery community participation. Activity volume and newcomer influx remained essentially unchanged following quarantines, suggesting that visibility reduction alone may not be sufficient to drive users toward recovery.
Second, we find banning fringe communities led to a marked increase in participation across all considered outcomes. These results suggest that hard moderation can act as a turning point, encouraging former members of fringe communities to seek support and begin the process of deradicalization (\textbf{RQ2}).
Third, our analysis suggests real-world events boosted participation in the recovery community, but the effects observed were smaller than those observed following bans (\textbf{RQ3}).

The robustness checks conducted across toxicity analysis, LLM moderation, manual inspection, control subreddit comparison, and deleted content analysis consistently support our conclusion: (hard) moderation interventions targeting fringe communities on Reddit led to increased participation in recovery communities like r/exredpill. This increased participation does not appear to be driven by negative or toxic behavior butreflective of genuine engagement with recovery processes.

\xhdr{Relation to existing social science theories} We discuss our findings in light of two prominent social science theories: the Role Exit Theory \cite{ebaugh1988becoming} and  \citet{aho1988out}'s Defection Model.
\citet{ebaugh1988becoming} theorizes the presence of turning points, events that lead to someone exiting a role. Our results indicate that bans, but not quarantines, may be understood as ``turning points.'' 
Also, in light of the social exit theory, we argue that recovery communities may help users redefine their identity (or, in the lingo of the theory, creating the ``ex-role'').
 \citet{aho1988out} theorizes deradicalization as a consequence of a change in the push and pull factors.
 For example, relationships with other people in a radical group could ``pull'' individuals toward the hate group, whereas relationships with minorities targeted by the group could ``push'' them away from it.
 In that context, this study analyzes the force of banning and quarantining as ``push factors,'' finding that banning seems enough to `flip the scale' for many individuals, whereas quarantining is not.
 
But why are community-wide bans impactful, whereas quarantines are not? We hypothesize that bans may sever social ties within the community~\cite{horta2021platform}, increasing the likelihood of users encountering alternative beliefs and counter-narratives within recovery communities. On the other hand, quarantines may merely isolate the community~\cite{chandrasekharan2022quarantined}, limiting opportunities for self-reflection or exposure to counter-narratives.

\xhdr{Implications} The results of this study highlight the potential for platform-based moderation to facilitate positive outcomes beyond merely reducing harmful activity ~\cite{chandrasekharan2017you,chandrasekharan2022quarantined}. Specifically, banning fringe communities appears to have the unintended but beneficial effect of driving a small but meaningful proportion of users to recovery communities, potentially initiating their journey toward deradicalization. Notably, we find that 4.7\% of all users with at least five posts in one of the moderated communities posted in the recovery community. Additionally, 78.3\% of the users who posted in the recovery community during this period continued to engage actively, averaging 12.7 posts over the following 120 days. This highlights the benefits of interventions that provide easier access to recovery communities or exposure to counter-narratives tailored to specific cohorts of users.
In light of this, we argue that platforms should consider how their moderation strategies, particularly bans, can be refined to guide users away from harmful ideologies and toward supportive environments. As platforms navigate the challenge of balancing free speech with user safety, these findings suggest that combining hard moderation strategies with targeted support mechanisms may offer a more comprehensive and effective approach to fostering recovery.

\xhdr{Broader Impact} 
While our findings suggest that platform moderation, particularly banning, may support deradicalization efforts, they also raise ethical questions about the potential consequences of deplatforming. Restricting users' ability to engage with certain content may lead to migration toward more radical and unregulated spaces~\cite{horta2021platform}, where extremism may further intensify. Platforms must, therefore, balance the benefits of sanctions with the risk of pushing users to more harmful environments.


\xhdr{Limitations and Future Work} While we took steps to mitigate potential confounders, such as analyzing real-world events, unobserved factors may still influence our results. Future studies could address these issues by incorporating more detailed user activity data, including passive engagement, and exploring the effects of moderation across platforms with different community structures. An important avenue for future research is to assess the long-term efficacy of recovery community participation after moderation actions. It remains unclear whether users who join recovery communities remain active, undergo genuine deradicalization, or eventually regress to their previous beliefs. Longitudinal studies focusing on user retention and shifts in ideological content would provide valuable insights into the sustainability of the recovery process. Last but not least, future work could focus on platforms different from Reddit, or fringe communities other than those within the Manosphere.

Last, we stress that, while we do find evidence that content moderation interventions may act as catalysts for deradicalization, our estimates represent a lower bound for three key reasons. First, our study captures only active contributors to r/exredpill, not passive participants (lurkers), meaning that even more users may have moved away from fringe communities. Second, we only consider users who engaged with recovery communities, which is likely only a fraction of users moving away from fringe communities.
Third, individuals redefining their identities may adopt new usernames, which our methodology cannot track.
We argue that these limitations do not decrease the importance of our findings, as a lower bound can still help us understand the consequences of content moderation interventions.

\bibliography{aaai25}


\subsection*{Paper Checklist}

\begin{enumerate}
\item For most authors...
\begin{enumerate}
    \item  Would answering this research question advance science without violating social contracts, such as violating privacy norms, perpetuating unfair profiling, exacerbating the socio-economic divide, or implying disrespect to societies or cultures?
    \answerYes{Yes}
  \item Do your main claims in the abstract and introduction accurately reflect the paper's contributions and scope?
        \answerYes{Yes}
   \item Do you clarify how the proposed methodological approach is appropriate for the claims made? 
        \answerYes{Yes}

   \item Do you clarify what are possible artifacts in the data used, given population-specific distributions?
        \answerYes{Yes}

  \item Did you describe the limitations of your work?
        \answerYes{Yes}

  \item Did you discuss any potential negative societal impacts of your work?
        \answerYes{Yes}

      \item Did you discuss any potential misuse of your work?
    \answerNo{No}
    \item Did you describe steps taken to prevent or mitigate potential negative outcomes of the research, such as data and model documentation, data anonymization, responsible release, access control, and the reproducibility of findings?
        \answerYes{Yes}
  \item Have you read the ethics review guidelines and ensured that your paper conforms to them?
        \answerYes{Yes}
\end{enumerate}

\item Additionally, if your study involves hypotheses testing...
\begin{enumerate}
  \item Did you clearly state the assumptions underlying all theoretical results?
        \answerYes{Yes}
  \item Have you provided justifications for all theoretical results?
        \answerYes{Yes}
  \item Did you discuss competing hypotheses or theories that might challenge or complement your theoretical results?
    \answerNA{NA}
  \item Have you considered alternative mechanisms or explanations that might account for the same outcomes observed in your study?
        \answerYes{Yes}
  \item Did you address potential biases or limitations in your theoretical framework?
        \answerYes{Yes}
  \item Have you related your theoretical results to the existing literature in social science?
        \answerYes{Yes}
  \item Did you discuss the implications of your theoretical results for policy, practice, or further research in the social science domain?
        \answerYes{Yes}
\end{enumerate}

\item Additionally, if you are including theoretical proofs...
\begin{enumerate}
  \item Did you state the full set of assumptions of all theoretical results?
    \answerNA{NA}
	\item Did you include complete proofs of all theoretical results?
    \answerNA{NA}
\end{enumerate}

\item Additionally, if you ran machine learning experiments...
\begin{enumerate}
  \item Did you include the code, data, and instructions needed to reproduce the main experimental results (either in the supplemental material or as a URL)?
    \answerNA{NA}
  \item Did you specify all the training details (e.g., data splits, hyperparameters, how they were chosen)?
    \answerNA{NA}
     \item Did you report error bars (e.g., with respect to the random seed after running experiments multiple times)?
    \answerNA{NA}
	\item Did you include the total amount of compute and the type of resources used (e.g., type of GPUs, internal cluster, or cloud provider)?
    \answerNA{NA}
     \item Do you justify how the proposed evaluation is sufficient and appropriate to the claims made? 
    \answerNA{NA}
     \item Do you discuss what is ``the cost`` of misclassification and fault (in)tolerance?
    \answerNA{NA}
  
\end{enumerate}

\item Additionally, if you are using existing assets (e.g., code, data, models) or curating/releasing new assets...
\begin{enumerate}
  \item If your work uses existing assets, did you cite the creators?
    \answerYes{Yes}
  \item Did you mention the license of the assets?
    \answerNo{No, given that the license is ill-defined.}
  \item Did you include any new assets in the supplemental material or as a URL?
\answerNo{No}
  \item Did you discuss whether and how consent was obtained from people whose data you're using/curating?
\answerYes{Yes}
  \item Did you discuss whether the data you are using/curating contains personally identifiable information or offensive content?
\answerYes{Yes}
\item If you are curating or releasing new datasets, did you discuss how you intend to make your datasets FAIR?
\answerNA{NA}
\item If you are curating or releasing new datasets, did you create a Datasheet for the Dataset? 
\answerNA{NA}
\end{enumerate}

\item Additionally, if you used crowdsourcing or conducted research with human subjects...
\begin{enumerate}
  \item Did you include the full text of instructions given to participants and screenshots?
\answerNA{NA}
  \item Did you describe any potential participant risks, with mentions of Institutional Review Board (IRB) approvals?
\answerNA{NA}

  \item Did you include the estimated hourly wage paid to participants and the total amount spent on participant compensation?
\answerNA{NA}

   \item Did you discuss how data is stored, shared, and deidentified?
\answerNA{NA}

\end{enumerate}

\end{enumerate}

\clearpage

\appendix

\subsection*{Prompts}

\begin{table*}[t!]
\small
\centering
\begin{tabular}{l D{.}{.}{2.6} D{.}{.}{2.6} D{.}{.}{2.6} D{.}{.}{2.8}}
\toprule
\multicolumn{4}{c}{Activity Volume}\\
\midrule
& \multicolumn{2}{c}{Braincels+TheRedPill} & \multicolumn{2}{c}{MGTOW} \\
\midrule
 & \mc{Quarantine} & \mc{Banning} & \mc{Quarantine} & \mc{Banning}  \\
\midrule
$\beta_1$ & -0.03       & 0.02         & 0.15      & 0.02       \\
          & (0.494)     & 0.853      & (0.014)     & 0.853      \\
                 
$\beta_2$ & -17.4     & 24.06      & 0.91         & 86.49     \\
         & (0.06)     & (0.001)          & (0.881)      & (0.001)         \\
                
$\beta_3$  & 0.16        & 0.259        & -0.0004	    & -0.63   \\
          & (0.002)     & (0.182)      & (0.988)      & 0.001       \\
                    
\midrule
\multicolumn{4}{l}{\scriptsize{\emph{Others}}}\\
\midrule
(Intercept)              & 10.737      & 22.124        & 9.344       & 98.432     \\
                         & (0.002)     & (0.000)      & (0.000)      & (0.001)    \\
\midrule
R$^2$                    & 0.017       & 0.025        & 0.047        & 0.014    \\
\end{tabular}
\begin{tabular}{l D{.}{.}{2.6} D{.}{.}{2.6} D{.}{.}{2.6} D{.}{.}{2.8}}
\midrule
\multicolumn{4}{c}{Newcomers}\\
\midrule
$\beta_1$               & -0.004       & 3.788   & 0.002         & -0.216       \\
                        & (0.771)      & (0.306)       & (0.114)       & (0.623)       \\
$\beta_2$               & -2.12        & 5.30    & -0.69          & 11.74^{***}          \\
                        & (0.261)      & (0.001)       & (0.663)       & (0.001)       \\
$\beta_3$               & 0.03         & 0.04     & 0.04          & -0.219^{*}    \\
                        & (0.130)      & (0.042)       & (0.988)       & (0.089)       \\
\midrule
\multicolumn{4}{l}{\scriptsize{\emph{Others}}}\\
\midrule
(Intercept)                    & 4.281^{***}  & 5.621         & 3.222         &  19.439    \\
                               & (0.001)      & (0.002)       & (0.001)       & (0.001)       \\
\midrule
Adj. R$^2$                     & 0.034        & 0.038         & 0.033         & 0.033       \\
\bottomrule
\multicolumn{4}{l}{\scriptsize{$^{***}p<0.001$; $^{**}p<0.01$; $^{*}p<0.05$; $^{\cdot}p<0.1$}}
\end{tabular}

\begin{tabular}{l D{.}{.}{2.6} D{.}{.}{2.6} D{.}{.}{2.6} D{.}{.}{2.8}}
\midrule
\multicolumn{4}{c}{Migrants from Fringe Communities}\\
\midrule
$\beta_1$                        & 0.167        & 3.788         & 0.005         & -0.216      \\
                                 & (0.267)      & (0.306)       & (0.172)       & (0.623)     \\
$\beta_2$                        & -0.01        & 1.24          & -0.031        & 21.10 \\
                                 & (0.318)      & (0.003)       & (0.058)       & (0.001)     \\
$\beta_3$                        &  0.044       & -0.012        & 0.029         & 0.042       \\
                                 &  (0.020)     & (0.318)       & (0.094)       & (0.441)      \\
\midrule
\multicolumn{4}{l}{\scriptsize{\emph{Others}}}\\
\midrule
(Intercept)                    & 2.859        &  3.206        & 6.408         & 18.123    \\
                               & (0.007)      & (0.001)       & (0.002)       & (0.003)       \\
\midrule
Adj. R$^2$                     & 0.023        & 0.012         & 0.027         & 0.018       \\
\bottomrule
\multicolumn{4}{l}{\scriptsize{$^{***}p<0.001$; $^{**}p<0.01$; $^{*}p<0.05$; $^{\cdot}p<0.1$}}
\end{tabular}
\caption{Summary of results. The ITS coefficient estimates (top) activity volume, (mid) newcomers, (bottom) number of migrants. Coefficient estimates and p-values in parenthesis.}
\label{table:coefficients}
\end{table*}

\begin{table*}[]
\centering
\renewcommand{\arraystretch}{1.1}
\begin{tabular}{c|c|c:c:c:c}
\textbf{banning event} &\textbf{outcome variable} & \textbf{p-value} & \textbf{relative effect} & \textbf{95\% CI}           & \textbf{absolute effect (s.d.)} \\ \toprule
\multirow{3}{*}{\textbf{Braincels ban}} & activity volume  & 0.032   & 88.4\%           & {[}12.5\%, 179.2\%{]}               & 1241 (883)                      \\
                               & new users        & 0.004   & 174.3\%           & {[}10.4\%, 388.2\%{]} & 87.2 (41.07)           \\
                               & migrating users  & 0.001   & 94.6\%            & {[}35.4\%, 207.2\%{]} & 45.71 (10.32) \rule[-1.5ex]{0pt}{0pt} \\ \hdashline
\multirow{3}{*}{\textbf{MGTOW ban}}     & activity volume  & 0.001   & 64.5\%   & {[}45.6\%,91.3\%{]}   & 5161 (564)              \rule{0pt}{2.6ex}\\
                               & new users        & 0.001   & 31.7\%            & {[}17.7\%,48.2\%{]}   & 166.8 (32.76)           \\
                               & migrating users  & 0.006   & 22.8\%            & {[}3.3\%,56.9\%{]} & 20.73 (12.0)           \\ \bottomrule
\end{tabular}
\vspace{10pt}
\caption{BSTS results of bans effects on volume activity, newcomers, and migrating users within r/exredpill.}
\label{table:banning-results}
\end{table*}

\begin{table*}[]
\centering
\renewcommand{\arraystretch}{1.1}
\begin{tabular}{c|l|r:r:r:r}
\textbf{quarantine event} &\textbf{outcome variable} & \textbf{p-value} & \textbf{relative effect} & \textbf{95\% CI}           & \textbf{absolute effect (s.d.)} \\ \toprule
\multirow{4}{*}{\textbf{Braincels \& TRP\textsuperscript{1} quarantine}}
& activity volume  & 0.345   & 15.4\%           & {[}-34.3\%, 65.9\%{]}  & 42.1(138.9)           \\
& new users        & 0.167   & 15.3\%            & {[}-19.6\%, 59.2\%{]}  & 17.43 (21.69)          \\
& migrating users   & 0.045   & 24.6\%            & {[}4.7\%, 53.1\%{]} & 13.03 (12.66)           

\rule[-1.5ex]{0pt}{0pt} \\ \hdashline
\multirow{3}{*}{\textbf{MGTOW quarantine}}           
& activity volume  & 0.093   & 38.5\%            & {[}-9.3\%,73.6\%{]}   &   797(264.9) \rule{0pt}{2.6ex}\\
& new users        & 0.223   & 13.7\%           & {[}-29.2\%, 47.8\%{]}  & 58.76(82.5)           \\
& migrating users  & 0.468    & 3.4\%          & {[}-25.3\%, 48.4\%{]}  & 0.77 (20.05)          \\ \bottomrule
\end{tabular}
\vspace{10pt}
\caption{BSTS results of quarantine effects on volume activity, newcomers, and migrating users within r/exredpill.}
\label{table:quarantine-results}
\end{table*}

\begin{table*}[t]
\centering
\renewcommand{\arraystretch}{1.1}
\begin{tabular}{c|c|c:c:c:c}
\textbf{banning event} &\textbf{outcome variable} & \textbf{p-value} & \textbf{relative effect} & \textbf{95\% CI}           & \textbf{absolute effect (s.d.)} \\ \toprule
\multirow{3}{*}{\textbf{Braincels ban}} 
& activity volume  & 0.105   & 10.5\%           & {[}-5.3\%, 22.3\%{]}               & 124 (88.3)                      \\
& new users        & 0.122   & 15.3\%           & {[}-4.2\%, 38.2\%{]} & 8.72 (4.11)           \\
& migrating users  & 0.132   & 9.6\%            & {[}-3.4\%, 20.7\%{]} & 4.57 (1.03) \rule[-1.5ex]{0pt}{0pt} \\ \hdashline
\multirow{3}{*}{\textbf{MGTOW ban}}     
& activity volume  & 0.114   & 6.5\%   & {[}-4.6\%, 9.1\%{]}   & 516 (56.4)              \rule{0pt}{2.6ex}\\
& new users        & 0.125   & 3.7\%            & {[}-1.7\%, 4.8\%{]}   & 16.7 (3.27)           \\
& migrating users  & 0.148   & 2.8\%            & {[}-3.3\%, 5.6\%{]} & 2.07 (1.2)           \\ \bottomrule
\end{tabular}
\vspace{10pt}
\caption{Placebo: BSTS results of banning effects on volume activity, newcomers, and migrating users within r/exredpill.}
\label{table:banning-results-placebo}
\end{table*}

\begin{table*}[]
\centering
\renewcommand{\arraystretch}{1.1}
\begin{tabular}{c|l|r:r:r:r}
\textbf{quarantine event} &\textbf{outcome variable} & \textbf{p-value} & \textbf{relative effect} & \textbf{95\% CI}           & \textbf{absolute effect (s.d.)} \\ \toprule
\multirow{4}{*}{\textbf{Braincels \& TRP\textsuperscript{1} quarantine}}
& activity volume  & 0.345   & 1.5\%           & {[}-3.4\%, 6.5\%{]}  & 4.21 (13.89)           \\
& new users        & 0.367   & 1.3\%            & {[}-1.9\%, 5.9\%{]}  & 1.74 (2.17)          \\
& migrating users   & 0.412   & 2.4\%            & {[}-0.7\%, 5.3\%{]} & 1.30 (1.27)           
\rule[-1.5ex]{0pt}{0pt} \\ \hdashline
\multirow{3}{*}{\textbf{MGTOW quarantine}}           
& activity volume  & 0.393   & 3.5\%            & {[}-0.9\%, 7.3\%{]}   &   7.97 (26.49) \rule{0pt}{2.6ex}\\
& new users        & 0.423   & 1.3\%           & {[}-2.9\%, 4.7\%{]}  & 5.88 (8.25)           \\
& migrating users  & 0.468    & 0.4\%          & {[}-2.5\%, 4.8\%{]}  & 0.08 (2.01)          \\ \bottomrule
\end{tabular}
\vspace{10pt}
\caption{Placebo: BSTS results of quarantine effects on volume activity, newcomers, and migrating users within r/exredpill.}
\label{table:quarantine-results-placebo}
\end{table*}

\begin{verbatim}


You are tasked with determining whether a comment posted by a migrating user
on the r/exredpill subreddit violates the community guidelines of r/exredpill. 

I will provide (1) the community guidelines and (2) the comment posted on r/exredpill.

Your answer must be formatted as follows:
1) "Answer: Yes" if the comment posted on r/exredpill violates the community guidelines

2) "Answer: No" if the comment does not violates the community guidelines


** Community Guidelines of r/exredpill ** : [...]

** Comment posted on r/exredpill **: [...]

Does the comment posted on r/exredpill violate the community guidelines?
\end{verbatim}

\begin{verbatim}
    
You are tasked with evaluating whether a comment posted 
by a user in r/exredpill aligns with the ideologies of fringe communities 
such as r/Braincels, r/MGTOW, and r/TheRedPill.
You will be provided descriptions of these ideologies and the comment to analyze.

Your answer must be formatted as follows:
1) "Answer: Yes" if the comment posted on r/exredpill violates the community guidelines

2) "Answer: No" if the comment does not violates the community guidelines

** Description of incels ideology  ** : [...]
** Description of MGTOW ideology  ** : [...]
** Description of TheRedPill ideology  ** : [...]


** Comment posted on r/exredpill **: [...]


Is the comment posted on r/exredpill aligned with the description of incels,

mgtow, theredpill ideology?

\end{verbatim}

\end{document}